\documentclass{svjour2}
\usepackage{graphicx}

\journalname{Gen. Relativ. Gravit.}
\begin{document}

\title{An Inhomogeneous Model Universe Behaving Homogeneously
\thanks{Paper in honor of Bahram Mashhoon's 60th birthday.}}
\author{Sh. Khosravi \and E. Kourkchi \and \\ R. Mansouri \and Y. Akrami }

\institute{Sh. Khosravi \at
              Physics Dept., Faculty of Science, Tarbiat Mo'alem
              University, Tehran, Iran. \\
              Institute for Studies in Theoretical Physics and
              Mathematics (IPM), Tehran, Iran.
              \email{khosravi@ipm.ir}
           \and
           E. Kourkchi \at
              Department of Physics, Sharif University of
              Technology, Tehran, Iran. \and R. Mansouri \at Department of
              Physics, Sharif University of Technology, Tehran, Iran. \\
              Institute for Studies in Theoretical Physics and
              Mathematics (IPM), Tehran, Iran.
           \and
           Y. Akrami \at
              Department of Physics, Stockholm University,
              AlbaNova University Center, Stockholm, Sweden.}

\date{Received: date / Accepted: date}

\maketitle

\begin{abstract}
We present a new model universe based on the junction of FRW to flat
Lemaitre-Tolman-Bondi (LTB) solutions of Einstein equations along
our past light cone, bringing structures within the FRW models. The
model is assumed globally to be homogeneous, i.e. the cosmological
principle is valid. Local inhomogeneities within the past light cone
are modeled as a flat LTB, whereas those outside the light cone are
assumed to be smoothed out and represented by a FRW model. The model
is singularity free, always FRW far from the observer along the past
light cone, gives way to a different luminosity distance relation as
for the CDM/FRW models, a negative deceleration parameter near the
observer, and correct linear and non-linear density contrast. As a
whole, the model behaves like a FRW model on the past light cone
with a special behavior of the scale factor, Hubble and deceleration
parameter, mimicking dark energy.
\keywords{Inhomogeneous Models
\and Dark Energy}

\end{abstract}

\section{Introduction}
\label{intro} Already in 1982, long before the detection of
acceleration of the universe and suggestions in favor of
inhomogeneous models as explanation for this novel cosmic effect,
Bahram published a work in collaboration with Hossein Partovi
\cite{Bahram84} in which they developed a model with radial
inhomogeneities as a first step toward a framework for comparison
with observations. The paper did not get much attention due to the
lack of observational evidence. The supernovae data indicating the
acceleration of the universe \cite{Riess98,Perl99} caused a revival
of the idea of an inhomogeneous universe. This time the authors
mainly considered Lemaitre-Tolman-Bondi (LTB) solution of the
Einstein equation as the model universe
\cite{Lemait33,Tolman34,Bondi47} and indicated a change in the
luminosity distance relation towards explaining the acceleration of
the universe due to the supernovae data (see \cite{Celer06} and
\cite{Mansouri05} for references). Now, most of the authors consider
the inhomogeneous solutions of the Einstein equations as an
alternative to FRW model of the universe (see
\cite{Biswas06},\cite{Celer07} and the references there). There is,
however, another way of understanding the inhomogeneity and the
interpretation of data on our past light cone. We know already from
many simple examples that cutting a manifold along a null
hypersurface and pasting it again after some warping we may get
gravitational shock or impulsive waves along the null hypersurface
even in an otherwise Minkowski spacetime. This has been nicely shown
in the paper by Penrose \cite{Penrose72}. The cut-and-paste method
is an elegant way to understand many astrophysical phenomena using
exact $C^0$ solutions of the Einstein equations. These solutions are
usually $C^3$ everywhere except across a null hypersurface of the
junction. Nevertheless, the inhomogeneous models of the universe as
{\it partial} explanation of dark energy, as Germans would say, are
just beginning to be {\it salonf\"ahig}. The task force for dark
energy \cite{DETF} does not consider inhomogeneities as a possible
explanation for dark energy. The first review or policy document in
which the inhomogeneity is explicitly mentioned as a possible way to
understand the acceleration of the universe
 seems to be ``The Newsletter of the Topical Group in Gravitation ..." \cite{APS}.\\

We will report here on a model universe based on a cut-and-paste
technology of matching an inhomogeneous model to a FRW model
universe across the past light cone of the observer. The philosophy
behind it is the following. After the decoupling era the universe,
being almost in a global FRW state, begin to develop as a matter
dominated universe witnessing a growth of structures. The structures
do not change the global homogeneous character of the universe, but
locally the inhomogeneities begin to grow. Any local observer, in a
globally Copernican view, will feel the effects of inhomogeneities
in the course of time. Needless to say that the inhomogeneities
outside his past light cone has no such effect and he may always
assume that everything outside his past light cone is effectively
homogeneous and is to be considered as a FRW model. On the contrary,
the inhomogeneities in his vicinity within the light cone, which
grows in time, may have direct cosmological consequences on his
observations. These effects have to be increasing with decrease of
the redshift on the past light cone. Therefore, we are faced with a
model universe in which the past light cone of the observer is a
boundary between a FRW model universe and an inhomogeneous model
which goes to a FRW with increasing $z$, to match the requirement of
almost FRW everywhere before the onset of nonlinearities in the
structures. For simplicity, we assume a LTB metric for the
inhomogeneous part. The matching should be such that no shear is
produced on the light cone as required for a simple boundary of
observation. We will see that as a matching condition the light cone
behaves as if we are in a FRW homogeneous model, but somehow
dilated, mimicking a dark energy not existent in either FRW or LTB
solutions on either side of the light cone. We have called this
model Structured FRW (SFRW), as opposed to the familiar homogeneous
FRW model of the
universe \cite{Mansouri05}. \\
The model we are considering is related to the Swiss Cheese model
but differs from it substantially. The Swiss cheese model of
cosmology, first suggested by Kantowski in 1969 \cite{Kant69},
studies the effects of local inhomogeneities in the FRW models on
the propagation of light through an otherwise homogeneous and
isotropic universe. Kantowski's model is constructed by taking the
Friedman model ($p = \Lambda = 0$), randomly removing co-moving
spheres from the dust, and placing Schwarzschild masses at the
center of the holes. The remaining Friedman dust is interpreted as
dust between clumps, and the point masses are interpreted as
inhomogeneities.
\\The paper has essentially two parts. In part one, we study in detail
the flat LTB solution of Einstein equations, its singularities, and
the bang time. This is just to get more insight how we are going to
use the LTB which differs substantially from its usage in cosmology
in the recent literature. This is done in section 2 where we have
written down the necessary formulas to define the LTB metric and
their corresponding Einstein equations, the bang time and
singularities, appropriate choice of the bang time, the past light
cone, and the place of singularities within it. Part one may be
skipped by those familiar with the LTB solution of the Einstein
equations, except for the numerical results showing the place of
singularities. In part two we define the model and study its
cosmological consequences. Section 3 is devoted to the explicit
definition of our SFRW model universe. In section 4 we look at the
cosmological consequences of our model, the age of the universe,
luminosity distance, deceleration parameter, and the density
contrast, followed by a section on conclusions.

\section{Flat LTB solution and related cosmological quantities}
\label{sec:1}
We constrain ourselves to the so-called flat or
marginally bound LTB models. These are solutions of Einstein
equations described by the metric

\begin{eqnarray}\label{metric}
ds^{2} = -c^2dt^{2} + R'^{\,2}\,dr^{2} +
R^{2}(r,t)(d\theta^{2}+\sin^{2} \theta d\phi^{2}),
\end{eqnarray}
in which overdot and prime (will thereafter) denote partial
differentiation with respect to $t$ and $r$, respectively. The
corresponding Einstein equations turn out to be
\begin{eqnarray}\label{field}
\dot{R}^{2}(r,t) &=& \frac{2GM(r)}{R} ,\\
4\pi\rho(r,t) &=& \frac{M'(r)}{R^2 R'}.
\end{eqnarray}
The density $\rho (r,t)$ is in general an arbitrary function of $r$
and $t$, and the integration time-independent function $M(r)$ is
defined as
\begin{equation}\label{density}
M(r)\equiv 4\pi \int^{R(r,t)}_{0}\rho(r,t)R^{2}dR
     = \frac{4 \pi}{3} \overline{\rho}(r,t) R^3,
\end{equation}
where $\overline {\rho}$, as a function of $r$ and $t$, is the
average density up to the radius $R(r,t)$. The metric (\ref{metric})
can also be written in a form similar to the Robertson-Walker
metric. The following definition
\begin{equation}
a(t, r) \equiv \frac{R(t, r)}{r}\,\, ,
\end{equation}
 brings the metric into the
form
\begin{equation}
ds^2 = -c^2dt^2 + a^2\left[\left(1 + \frac{a' r}{a}\right)^2 dr^2+
r^2d\Omega^2 \right].
\end{equation}
For a homogeneous universe $a$ doesn't depend on $r$ and we get the
familiar Robertson-Walker metric. The corresponding field equations
can be written in the following familiar form
\begin{equation}
\big (\frac{\dot a}{a} \big)^2 =  {G\over 3}\frac{\rho_c(r)}{a^3},
\end{equation}
where we have introduced $\rho_c(r) \equiv \frac{6M(r)}{r^3}$
indicating a quasi comoving $r$-dependent density. These are very
similar to the familiar Friedman equations, except for the
$r$-dependence of different quantities. The solutions to the field
equations can be written in the form
\begin{eqnarray}
R(r,t) &=&  \left[\frac{9GM(r)}{2}\right]^{1\over 3}
\left[t- t_n(r)\right]^{2\over3}, \nonumber  \\
a(r,t) &=& \left[\frac{3}{4}\, G\rho_c(r)\right]^{1\over
3}(t-t_n(r))^{2\over3}.
\end{eqnarray}
Now we choose the coordinate $r$ such that
\begin{equation}\label{r}
M(r) = \frac{4\pi}{3} \rho_c r^3,
\end{equation}
where $\rho_c$ is a constant \cite{Biswas06}. To adapt this metric
to observational data we need to know the backward light cone, the
luminosity distance, the corresponding Hubble parameter, the
deceleration parameter, the jerk, and the equation of state
parameter $w$. We start with the radial light rays. The null
geodesic corresponding to radially inward rays is given by
\begin{equation}\label{dt}
cdt = - R'(r, t)dr.
\end{equation}
Assuming the redshift $z$ as the parameter along the past light
cone, we obtain \cite{Celer00}
\begin{equation}\label{rz}
\frac{dr}{dz} = \frac{c}{(1+z)\dot R'(r,t(r))},
\end{equation}
and
\begin{equation}\label{tz}
\frac{dt}{dz} = \frac{-R'(r, t(r))}{(1+z)\dot R'(r,t(r))},
\end{equation}
where $t(r)$ is evaluated along the rays moving radially inward
according to (\ref{dt}). The luminosity distance is then given by
\begin{equation}\label{ld}
D_L(z) = (1+z)^2 R.
\end{equation}
Definition of the Hubble function is not without ambiguity in the
LTB models. The reason is $r$ and $t$ dependence of the ``scale
factor $a$" \cite{Moffat05,Moffat06}. Depending on the use of the
metric coefficients $R$ or $R'$ we may define $H_R = \frac{\dot
R}{R}\,$ or $H_{R'} = \frac{\dot R'}{R'}$ as the Hubble parameter,
which are different in general. We may also define a Hubble
parameter as the expansion rate along the light cone
\cite{Vander06}. It gives us a quantity which is easy to compare
with the observations. Once we have the LTB-luminosity distance from
(\ref{ld}), we interpret it as a distance in an effective FRW
universe. Assuming the relation
\begin{equation}
D_L(z) =(1+z)\int_{0}^{z} \frac{dz^{'}}{H(z^{'})}\,\, ,
\end{equation}
we invert it to define
\begin{equation}\label{h}
H(z) = \left[\frac{d}{dz}\frac{D_L(z)}{1+z}\right]^{-1}.
\end{equation}
Note that in our structured FRW model this definition corresponds to
the Hubble parameter of the FRW background outside the past light
cone. Therefore, in general, we may have three different definitions
of the Hubble parameter. It has been shown, however, that in the
case of glued LTB to FRW along a null hypersurface the three
definitions coincide \cite{khakshour07}. A numerical comparison is
given in section 4.4 and Fig.14 \\
The associated deceleration parameter is then defined as
\begin{equation}\label{q}
q(z) = -1 +\left [\frac{1+z}{H(z)}\right]\frac{dH(z)}{dz},
\end{equation}
and the effective state parameter
\begin{equation}\label{w}
w \equiv \frac{2}{3}[q(z)- {1/2}] =
\frac{2(1+z)}{3}\frac{d}{dz}\ln\left[
\frac{H(z)}{(1+z)^{3/2}}\right]\,\, .
\end{equation}
In addition, we may define the jerk as in a FRW universe
\cite{Mobasher04,Sahni03}:
\begin{equation}\label{j}
j=\frac{\dot{\ddot{a}}}{a}(\frac{\dot{a}}{a})^{-3}=\frac{(a^2H^2)''}{2H^2},
\end{equation}
where prime means derivative with respect to the argument $a$, and
substitute an effective scale factor to obtain the corresponding
value for our LTB based model. In this way we have a simple way to
compare any LTB model with an effective FRW universe and also with
the observational data. We may consider jerk as an alternative to
$w$ for parametrization of the dark energy. In our case it is to be
considered as another way of interpreting inhomogeneities in terms
of an effective FRW cosmology. \\
We will see in section 3 that the effective FRW cosmological
parameters defined here are the exact FRW quantities for our SFRW
model on the light cone and that different Hubble parameter
definitions for the LTB metric coincide with each other and are the
same as the Hubble parameter of the background FRW on the light
cone.

\subsection{Bang time and singularities}
\label{sec:2} The singularities of LTB metric, which are more
sophisticated than in the case of the Robertson-Walker metric, have
been discussed extensively in the literature (see for example
\cite{Vander06}, \cite{Christo84}, \cite{Newman86}). Vanishing of
each of the metric functions and its derivatives $R, R', \dot R,
\ddot R, \dot R',\newline R''$ may lead to different singularities.
In a general LTB metric there is another singularity, the event
horizon, related to zero of $1+ E$, where $E(r)$ is the energy
function of the LTB metric, absent in our flat LTB case. Vanishing
of $R''(t, r=0)$ leads to the so-called central weak singularity in
LTB models studied in \cite{Vander06}. For this central singularity
in a flat LTB model to be absent one must have \cite{Vander06}
\begin{equation}\label{nonsing}
t_{n}'(0)= 0.
\end{equation}
In selecting a bang time we will take this into consideration and
will avoid those violating this condition. The place of the other
singularities, irrespective of its cosmological significance, are summarized as follows:\\
\begin{eqnarray}
& &R',R'',\dot{R},\dot{R'} = \infty \hspace{.5cm}\longrightarrow\hspace{.5cm} t=t_n\,\, , \nonumber\\
& &R =0 \hspace{2.3cm}\longrightarrow\hspace{.5cm} t=t_n\,\, ,  \nonumber\\
& &R' = 0 \hspace{2.2cm}\longrightarrow\hspace{.5cm} t=t_n+\frac{2}{3}rt'_n\,\, ,\\
& &\dot{R'}= 0 \hspace{2.2cm}\longrightarrow\hspace{.5cm} t=t_n-\frac{1}{3}rt'_n\,\, , \nonumber\\
& &R'' = 0 \hspace{2.1cm}\longrightarrow\hspace{.5cm}
t=\frac{3rt_nt''_n+6t_nt'_n -r{t'_n}^2}{6t'_n+3rt''_n}\nonumber.
\end{eqnarray}
The sign of $t'_n$, as it may be seen from the above relations,
plays an important role in the discussion of the singularities.
Taking note of these singularities, we will construct our model such
that none of the singularities is on the light cone and all of them
are far from the region where the LTB metric is effective.
\subsection{Choice of the bang time}

According to SFRW model of the universe, the cosmological principle
is valid. The universe from the point of view of each observer looks
the same: inhomogeneous locally but homogeneous far from the
observer and everywhere outside its past light cone. Of course,
outside the light cone we have everywhere the local inhomogeneity
which may affect us in future, but it is assumed to be smoothed out
to a FRW homogeneous universe, as it is always assumed in the
familiar FRW model of the universe. The difference is on the light
cone and in the vicinity of the observational point, where we do not
want to assume the smearing out of inhomogeneities and would like to
model it using a flat LTB solution of the Einstein equations, being
glued to the outside flat FRW metric \cite{khakshour07}. \\
To formalize this requirement, we fabricate a bang time so that for
$r$ greater than a fiducial distance, say $nL$ of the order of
magnitude 100 MPc, the metric goes over to FRW, i.e. $t_n
\rightarrow const$ for $r \gg nL$. Here we assume $L =100$ MPc and
leave $n$ to be determined by the observation. For numerical
calculation, we therefore define a dimension-less co-moving distance
$r' = \frac{r}{nL}$. For the sake of simplicity we rename from now
on $r'$ as $r$. Our coordinate $r$ is now dimensionless and scaled
by the inhomogeneity scale $nL$. But note that we may use $r$ for
the comoving coordinate or the scaled version of it interchangeably
according to the context it is used! Now, for the bang time to have
the desired property, we may write it in the following general form:
\begin{equation}
t_n = \frac{\alpha}{p(r) + 1 + q({1\over r})},
\end{equation}
where $p$ and $q$ are polynomials in their arguments having no
constant and linear term. The time factor $\alpha$ is another
constant of the model in addition to inhomogeneity parameter $n$. It
is obvious that for $r\ll 1$ the bang time approaches a constant, in
fact zero for $q \neq 0$, and for $r\gg 1$ it approaches a constant,
zero again for $p\neq 0$, meaning that for large $r$ we have
effectively FRW metric again. Note that large $r$ will corresponds
in our case to redshifts bigger than 1. To see the cosmological
effects of polynomials like $p(r)$ we restrict it to powers of up to
$r^4$. On the other hand, it is easily seen that large powers of $1
\over r$ have not much effect and we can ignore them within the
scope of this paper. We, therefore, ignore them altogether to
concentrate on the most significant cases and effects. Now, consider
just the following two cases:
\begin{eqnarray}
t_{n1} &=& \frac{\alpha}{r^2 + 1}\\
t_n &=& \frac{\alpha}{r^4 + r^2 + 1}\, .
\end{eqnarray}
These bang times are just two examples of non-singular cases we will
concentrate on. In fact, there are many choices leading to almost
FRW for large z, as we have tested numerically. They are, however,
typically mal-behaved as far as singularities, density contrast, or
deceleration is concerned, as it is the case with $t_n =
\frac{\alpha r^2}{1 + r^2}$ suggested in \cite{Vander06}. We will
see that the bang time suggested in this paper is free of any
singularity on the light cone and reflects a reasonable behavior of
the luminosity
distance and the deceleration. \\
 We have plotted the behavior of $t_n$ and $t_{n1}$ as a function of $r$ in Fig.
\ref{fig1}. Both bang time functions decay very rapidly to zero,
which is equivalent to rapid decay of the LTB metric to FRW.
Therefore, it reflects the desired feature of SFRW. The $t_{n1}$,
however, leads to singular behavior at the light cone for $z$-vlaues
larger than $600$. That is why we prefer to work with $t_n$. In both
cases, however, as it is shown in the following sections, the
difference between LTB and FRW is almost negligible for redshifts $z
> 1$.

\subsection{Past light cone and the place of singularities in the flat LTB model}

There are many papers dealing with singularities of the LTB metric.
It is claimed that LTB metrics have either a weak singularity at the
origin or do not lead to an acceleration \cite{Flan05}. Here we
report on the singularities of $t_n$ and $t_{n1}$ as defined above,
just as two examples of bang time being non-singular at the origin.
The question of acceleration is dealt with in the next section. In
Figs. \ref{singul02} and \ref{singul04} we have plotted our past
light cone and different singularities within it for $t_n$ and
$t_{n1}$. The singular points of $R''$ are also sketched in the
mentioned figures. This curve intersects the past light cone on a
point which corresponds to a local maximum of $R'$ as can be seen
from Figs. \ref{rp1} and \ref{rp2}, and none of the metric
invariants, including that of Kretschman, have a singularity at this
event. Therefore, we notice that all singularities are within the
light cone, well before the time $t = \alpha$, and well outside the
vicinity of the light cone in the region where we have effectively
FRW again.
\begin{figure}
\centering
\includegraphics[angle=-90, scale=.4]{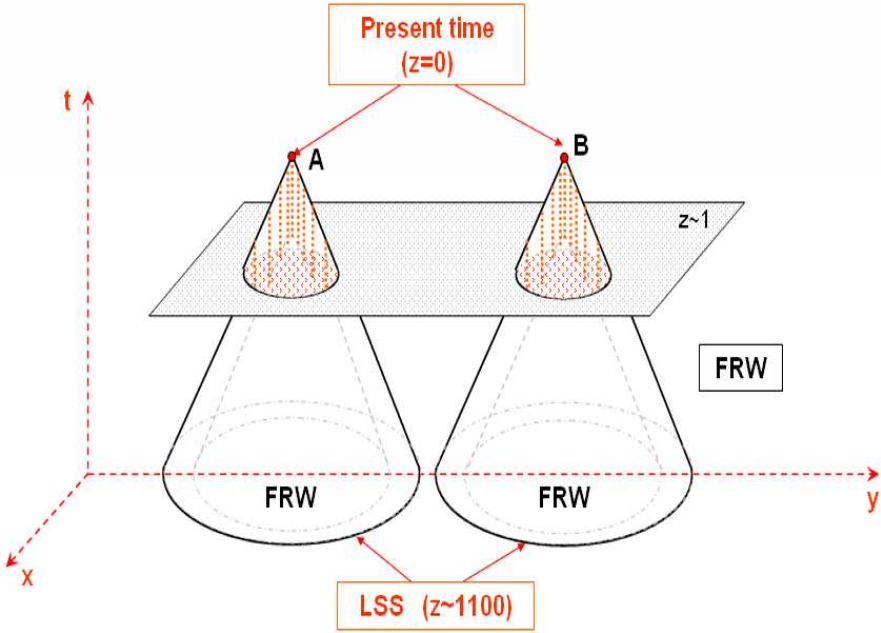}
\caption{Observers A and B have the same picture of the universe:
affected by the local inhomogeneities in their vicinity ( above the
$z=1$ plane); Outside their past light cone and far from the
observer points A and B, the universe seems homogeneous.}
\label{sfrwmodel}
\end{figure}

\section{Definition of the model: Structured FRW Universe}

We are used to the Swiss cheese model as a random distribution of
over- and under-dense regions in a constant time slice of the
space-time, a constant time picture that does not take into account
the realities of the observations along the past light cone and the
possible evolutionary effects. We intend to modify this picture
taking into account that all our observations are along the past
light cone, and these observations are not affected by the events
outside it.\\
Let us accept the cosmological principle according to which all
observers have the same picture of the universe. As the structures
are effectively influencing the universe after the last scattering
surface, our goal is to model the matter dominated and pressure-less
universe. The universe before the last scattering surface is
radiation dominated  and structure-less, i.e. it is represented by a
FRW metric. After the last scattering and with the growth of
structures, our model gradually deviates from the simple CDM/FRW
universe in the following way. We will use a combination of FRW and
LTB metrics connected to each other along the past light cone in a
specific manner. Contrary to the familiar concept of using LTB
metric to represent over- or under-density bubbles in the time
constant slices of the universe in many papers published in the last
years, we use it in a light cone adapted way. We will see that due
to our model construction the effect of LTB inhomogeneities vanishes
on the light cone and we encounter another effective FRW which
differs from the CDM/FRW, although we
start with a CDM/FRW in combination with a pressure-less LTB. \\
The universe is assumed to be globally FRW, but structured locally.
Hence, any observer sees the universe in the following way: outside
his past light cone, which is not observable to him, we assume the
universe to be FRW, i.e. local inhomogeneities outside the light
cone are assumed to be smoothed out and the effective metric is FRW.
Note that the only use of this assumption is the induced Hubble
constant on the light cone. Now, to implement the effect of the
local inhomogeneities in the spirit of SFRW model, we assume the
metric on the past light cone and its vicinity to be LTB. to define
the vicinity we note that the order of magnitude of inhomogeneities
is given by $nL$. Now at any $z-$value, the vicinity of the light
cone can be determined according to the size of horizon. In fact, we
are only concerned with the light cone behavior of the metric in
this paper. In a following paper we will define the vicinity more
precisely to study the averaging process of the Einstein equations
and answer the question of the backreaction of inhomogeneities, its
pitfalls, and how to make sense of it \cite{khosravi07}. At any
point outside this region the observer sees on the average, a FRW
homogeneous universe again. Although, FRW and LTB could in principle
be any of the three cases of open, flat, or close, it has been shown
in \cite{khakshour07} that the only meaningful matching of these two
spaces along a null boundary is a flat-flat case. In addition, due
to the cosmological preferences we
will assume both metrics to be flat (Fig. \ref{sfrwmodel}). \\
Note the philosophy behind this picture: The universe is everywhere
locally inhomogeneous, but homogeneous at large. Any observer,
according to the cosmological principle, has the same picture. Now,
he makes his model according to these rules: universe outside his
past light cone, i.e. regions of the universe that has not
influenced his past, is FRW. Locally, i.e. not far along the past
light cone or for small $z$-values which will turn out to be of the
order of $z = 1$, the observer sees a flat LTB on and within it. Far
on the light cone and within it the model tends to be a homogeneous
FRW universe again. This is of course in contrast to the usual
assumption in the recent literature
that the LTB represents bubbles of inhomogeneities. \\
Now, let us assume the bang time to be
\begin{equation}\label{bang}
t_n = \frac{\alpha}{r^4 + r^2 + 1},
\end{equation}
where $r$ is scaled to $nL$. We will alternatively use the comoving
coordinate $r$ in the scaled form or not, and the reader may simply
see from the context which one is meant. The constant $\alpha$ has
the dimension of time. We have, therefore, two model parameters $n$
and $\alpha$ to be determined by observation. Fig. \ref{tnRr} shows
$t_n$ as a function of the redshift $z$, using (\ref{rz}). We have
plotted $r$ and $R$ as a function of $z$ in the same diagram to see
 the coordinate, or redshift for which our LTB metric is essentially
FRW. Obviously, the value $z \approx 1$, corresponding to $r \approx
10$ and $t \approx 2\alpha$, is the boundary of transition from LTB
to FRW. The inhomogeneity scale $r \approx 10$, or roughly 1000 Mpc,
corresponds to the physical length $\approx 1$ Mpc at the time of
the last scattering for $z\approx 1100$. Although the transition
point is at a relatively small $z$, we can follow the effect of
inhomogeneity up to the last scattering surface at $z \approx 1100$.
The bang time is zero almost everywhere, except in our vicinity.
Therefore, if the inhomogeneity has any effect, it must show up at
our time, in accordance with the cosmological coincidence.

Now, for this bang time (\ref{bang}) we have
\begin{equation}
t'_n|_{r = 0} = 0.
\end{equation}
Therefore, we do not expect any weak singularity at the origin
\cite{Vander06}. In fact, for the LTB domain with this bang time, we
have no singularity at all, as it is shown in Fig. \ref{singul02}.
Vanishing of $R''$ at the light cone and its vicinity indicates a
maximum of $R'$ for $t= const.$ which reflects a feature of
expanding layers. No invariant of the metric has a singularity
within the domain of our interest. The singularity $t = 3t_n$ which
appears in the Kretschman invariant is well outside the region of
our interest as shown in Fig. \ref{singul02}. We have also plotted
in Fig. \ref{rp1}, $R'$ as a function of $r$ for some fixed values
of time to have a better understanding of variation of the LTB
metric function. Note the behavior of $R'$ near its maximum values
where $R''$ vanishes and compare it with Fig. \ref{singul02}. The
behavior of the LTB scale factor $R/r$ and the metric function $R'$,
for the sake of comparison the FRW scale factor for the CDM case
($\alpha = 0$) as a function of $z$, is depicted in Fig.
\ref{scalefactor}. The density as a function of redshift is also
plotted in Fig. \ref{rho1} and Fig. \ref{rho2} for different ranges
of $z$.

\section{Cosmological consequences of SFRW and observational data}

We are now in a position to look at different cosmological
consequences of the model. Angular distance and the luminosity
distance are given implicitly by the equations (\ref{rz} -
\ref{ld}). We have to integrate from $z = 0$ to the last scattering
surface. The definition of the last scattering surface is a delicate
problem. Its familiar value $z \approx 1100$ is based on a FRW
model. But our SFRW model tends to a FRW at early times. Therefore,
to match the SFRW at early times to a radiation dominated FRW we
have to have the same age for the universe at the last scattering
surface. This fixes the $z$-value in SFRW, which may differ from the
one in FRW. It turns out, that for our choice of the bang time the
difference is sensitive to the model parameters. Note, however, that
for many other choices of the bang time there is no meaningful
matching to the radiation dominated universe irrespective of any
choice of the parameters. For the range of parameters we are
considering, i.e. $0.1< n <3$ and $\alpha \approx 1$ ( in scale of
$10^{17} s$) , the difference is negligible and we can fix the last
scattering surface at $z \approx 1100$. Before going into its
detail, we have to fix another constant of the model, namely
$\rho_c$. It is determined by the value of the scale factor of the
universe at the last scattering surface. Taking it the same as that
of FRW at the age of $\approx 10^{13}~s$, we arrive at $\rho_c
\approx 10^{-30} g/cm^{3}$ or $10^{-37}$ in geometrical units.

\subsection{Angular distance and the age problem}

We have fixed now all the constants of the model. The last
scattering surface is defined now by $z= 1100$, corresponding to the
cosmic time $t \approx 10^{13} s$. Integrating the (\ref{tz}), we
end up with Fig. \ref{lightcone} for the light cone. For comparison,
we have also plotted the CDM/FRW light cone for the same redshift
interval. The age of the universe at the last scattering surface is
almost the same for both models. Therefore, the value of time for
$z= 0$ determines the age of the universe in SFRW. It turns out to
be $4/3$ of the age in FRW, i.e. $t = 3.9 \alpha = 15\times 10^{9}$
years. Therefore, there is no age problem in the model: the age of
the universe in our model is well above any estimation we have in
astrophysics and cosmology \cite{Carretta00} (see Fig.
\ref{lightcone}). The angular distance defined by $R(r,t)$ as a
function of $z$ is plotted in Fig. \ref{tnRr}.\\

\subsection{Luminosity distance}

The model is now completely determined. The last scattering surface,
having the same age in SFRW and FRW, corresponds to $z \approx
1100$. But as we have seen in the last section, the age of the
universe is $4/3$ of that in FRW. Now, we can integrate the
equations (\ref{rz} - \ref{ld}) to obtain the luminosity distance.
Fig. \ref{lumino} shows the luminosity distance as a
function of $z$.\\
For the sake of completeness and more insight in the models under
consideration, we have listed the results of likelihood analysis for
the luminosity distance of type Ia supernovae (SNe Ia - GOLD samples
\cite{Mobasher06}) given different forms of the bang time $t_n$ in
the table \ref{chi2} (
assuming $nL \simeq 100 Mpc$ and $\alpha=10^{17}~s$). \\

\begin{table}[h]
\begin{center}
\begin{tabular}{l c}
\hline
$p(r)$ & $\chi^2$ (for 182 SNe Ia)\\
\hline
$r^4+1$     &  255 \\
$r^4+r^2+1$     &  265 \\
$r^2+1$     &  226 \\
$r^2+1+r^{-1}$     &  234 \\
$r^2+r+1+r^{-1}+r^{-2}$     &  239 \\
$r^2+1+r^{-1}+r^{-2}$     &  251 \\
$r+1$     &  209 \\
$r+1+r^{-1}$     & 220  \\
$r+1+r^{-1}+r^{-2}$     &  226 \\
$0.0001r^4+0.001r^2+1$     &  286 \\
\hline
$CDM$ &  1700 \\
$\Lambda CDM~~(\Omega_\Lambda=0.7)$ & 393 \\
\hline
\end{tabular}
\caption{$\chi^2$ test for different choices of bang time. Bang time
is defined as $t_n=\frac{\alpha}{p(r)}$ and we put $n=1,
\alpha=10^{17}s, t_0=3.9 \times 10^{17}s $.} \label{chi2}
\end{center}
\end{table}

Different bang times $t_n$ having a wide range of free parameters
(i.e. $\alpha$ and $n$) enables us to get best fit luminosity
distances, and likelihoods as in the other models like LCDM. Of
course, we need to compare the results of our model with other
cosmological data to choose the most suitable bang time and the
acceptable range of its free parameters. At this stage, we found out
that $\alpha$ should be approximately $10^{17} s$ for any bang time
$t_n$. Looking at the density contrast, we can estimate the
inhomogeneity factor as well (see section 4.4). Table
\ref{parameter} shows the best parameters for different bang times
using SNe Ia data (GOLD samples):

\begin{table}[h]
\begin{center}
\begin{tabular}{lccc}
\hline
Bang time & $\alpha~( \times 10^{17} s)$    &   $n$  &   $\chi^2$ (for 182 SNe Ia)\\
\hline
$\alpha(r^2+1)^{-1}$ & 1.1  & 2.2 & 188 \\
$\alpha(r^4+1)^{-1}$ & 1.0 & 2.2 & 212 \\
$\alpha(r^4+r^2+1)^{-1}$ & 1.0 & 2.9 & 198 \\
\hline
\end{tabular}
\caption{Best parameter choice for different bang times.}
\label{parameter}
\end{center}
\end{table}

\subsection{Effective scale factor, Hubble parameter, and deceleration parameter}

There are different ways to define an effective scale factor for an
LTB metric. Both $R/r$ and $R'$ could serve as effective scale
factors, and correspondingly one may define $\frac{\dot R}{R}$ or
$\frac{\dot R'}{R'}$ as the Hubble parameter. Of course, the
definition (\ref{h}) we used to define the effective Hubble
parameter may also differ from the other two definitions. It is
interesting, however, to note that all three definitions coincide if
the LTB metric is matched exactly to FRW metric along the light
cone, as has been shown in \cite{khakshour07}. The requirement of
the exact matching leads to a definite bang time, which could be
calculated numerically. In fact, what we have done is the inverse
problem: we have been looking for a suitable bang time mimicking the
exact matching. What we have achieved is a bang time which leads
almost everywhere, for the redshift values $1100 < z < 0.03$, to the
equivalence of all three Hubble parameter, i.e. to an exact
matching. The actual problem of requiring exact matching and finding
out the bang time is subject of a future work.\\
Using relations (\ref{h}) and (\ref{q}), we have plotted in Fig.
\ref{hubble} the different Hubble parameters. For comparison, we
have also plotted $H-$values for CDM and LCDM models in the
homogeneous FRW universe models ($\alpha = 0$). Note the peculiar
$z-$dependence of the Hubble parameters for $z < 0.5$. There we
observe a decrease of the Hubble parameter up to a value of about
$50$ and increase again to the present value of about $70$. This
behavior may be partially due to the deviation of $H_R$ from $H =
H_{R'}$ \cite{Sarkar03}.\\
The effective deceleration parameter, as defined in (\ref{q}), is
plotted in Fig.\ref{deceler} for $t_{n1}$, used in our SFRW model.
It shows the effective deceleration for different $n-$values. For $z
\gg 1$ all models almost coincide and have $q= 1/2$. They, however,
differ from each other for small $z$. Negative deceleration is
almost typical for all of them. Fig.\ref{deceler2}
shows the deceleration parameter for $t_{n}$.\\

\subsection{Density contrast}

Let us now look at the density contrast within the LTB domain. At
any time, the density contrast within the LTB domain defined above
is given by
\begin{equation}\label{delta}
\delta = \frac{\delta_{max} - \delta_{min}}{\delta_{min}}.
\end{equation}
Calculation of $\delta$ is straightforward. We expect $\delta$ to be
of the order of magnitude $10^{-5}$ for $z \approx 1100$ and of the
order of magnitude 1 at the present time. Using $t_n$, there is a
rather wide range of $\alpha$ and $n$ satisfying the density
contrast conditions. Table \ref{tableden} shows the density contrast
for some of the parameter values.
\begin{table}
\begin{center}
\begin{tabular}{l c l}
\hline
$\alpha\, (\times 10^{17}s)$  &$n$&  \\
\hline
$<0.5$    &  --- & \textit{Not valid}\\
$1$     &  $n \sim 1$ & $\delta(z \approx 1100) \sim 10^{-7}$     \\
$1$     &  $2<n<3$ &  \textit{Valid for both conditions} \\
$1.5$     &  $1<n<4$ &  \textit{Valid for both conditions} \\
$2$     &  $3<n<4$ &  \textit{Valid for both conditions} \\
$3<\alpha<5$     &  $4<n<5$ &  \textit{Valid for both conditions} \\
\hline
\end{tabular}
\caption{Density contrast for $t_n$ and different parameter values.}
\label{tableden}
\end{center}
\end{table}

The above results for $t_n$, and those of the luminosity distance
($\alpha=10^{17}~s$), shows that the inhomogeneity factor $n$ should
be chosen in the range $1<n<3$. Note that the density contrast
should not be greater than the desired order of magnitudes. In the
opposite case, one can  compensate the lack of contrast by adding
some density perturbations as it is done in FRW models. In the case
of $t_n$ for $\alpha = 1$ the inhomogeneity factor $n$ has to be
$<1$ to achieve the correct density contrast at the last scattering
surface.

\section{Conclusion}

The Structured FRW model of the matter dominated universe we are
proposing is a singularity free model which incorporates the concept
of local inhomogeneities along the past light cone and is globally
FRW. Although locally we have used the LTB metric, the effective
model is a FRW type: along the light cone different Hubble
parameters coincide, except for small values of z, as if we have
just a FRW model. Therefore, we could consider this SFRW as a
modified CDM/FRW with novel features. This is in contrast to
familiar use of LTB metric to model local over- or under-dense
bubbles. Not only the age of the universe is increased, the
deceleration parameter has also an interesting behavior: it is
negative for small $z$-values and goes to $1/2$ for larger
redshifts. The density contrast changes from the order of magnitude
$10^{-5}$ at the time of the last scattering surface to $1$ for the
present time. Our model has two parameters $\alpha$ and $n$ both of
the order of magnitude 1. The bang time is defined in a way that
there is no central weak singularity present in the model; although
it is not unique, the difference between different bang times in the
class we have defined is marginal. We conclude that allowing for
local changes along the past light cone one may construct model
universes that behave Friedmanian but differs from the CDM model and
leads to local acceleration, at the same time increases the age of
the universe and reflects the growth of density contrast.

\begin{acknowledgements}
Authors wish to thank Sima Ghassemi for her valuable comments. R.M.
would also like to thank ISMO (Tehran) for partial support through
TWAS-IC funds.
\end{acknowledgements}

\bibliographystyle{spmpsci}
\bibliography{SFRWbib}
\newpage

\begin{figure}
\centering
\includegraphics[angle=0, scale=.8]{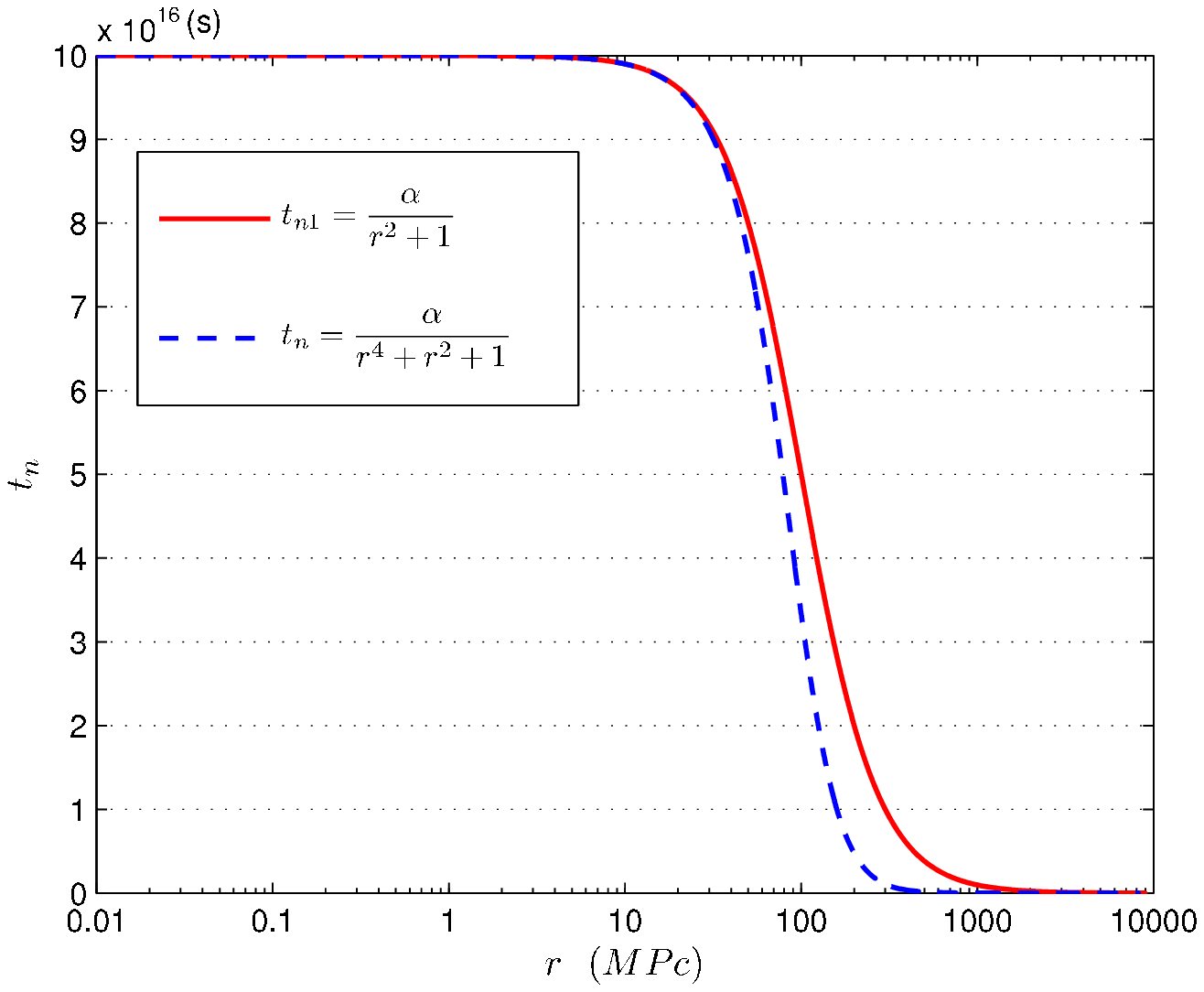}
\caption{ $t_n$ vs r using different functions. The blue curve for
$t_{n}=\frac{\alpha}{r^4+r^2+1}$ and the red curve for
$t_{n1}=\frac{\alpha}{r^2+1}$.} \label{fig1}
\end{figure}

\begin{figure}
\centering
\includegraphics[angle=0, scale=.8]{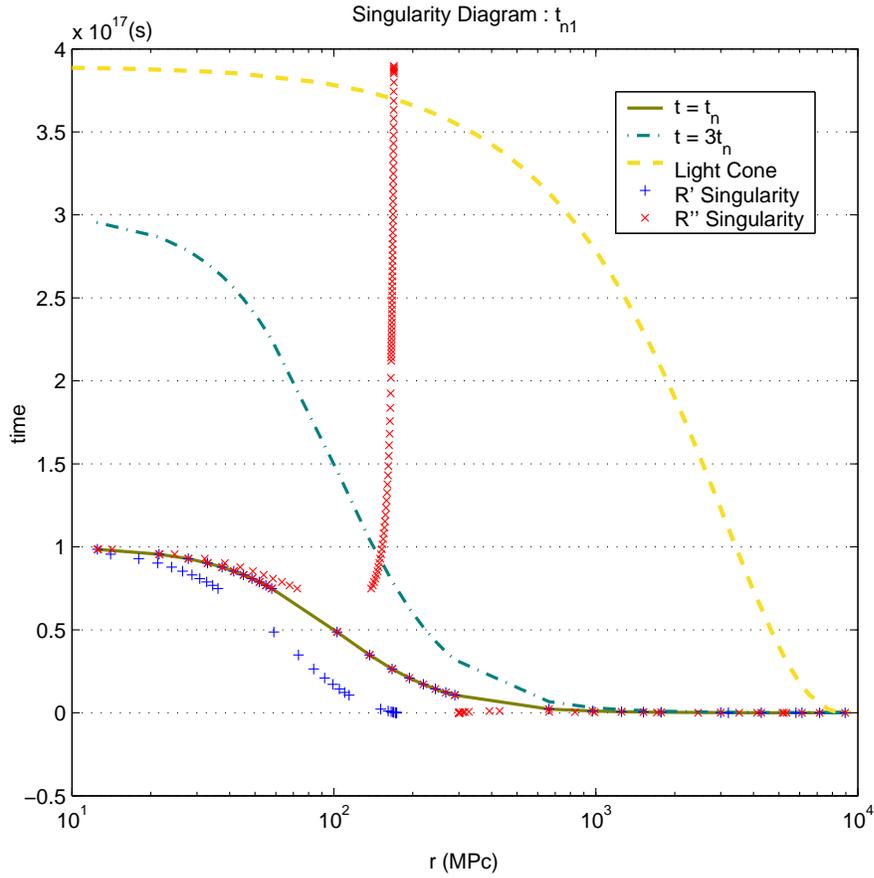}
\caption{ Light cone is shown (time vs $log_{10}(r)$) for
$t_{n1}=\frac{\alpha}{r^2+1}\,\, , (\alpha=10^{17} s, n=1)$. The
solid line curve shows the shell focussing area $R = 0$ at $t=t_n$.
On this curve both $R'$ and $R''$ are singular. Here, one can easily
see that one of these singular points lie on the light cone. This
singular point maybe avoided if we restrict the value of n to be
less than 0.2 or choose another bang time $t_{n}(r)$. The shell
focussing singularity crosses the light cone at $z-$values well
above the last scattering surface which is outside the range of
applicability of our model, although it may also be avoided using
suitable inhomogeneity parameter. The dash-dotted green curve
corresponds to Kretschman singularity ($t=3t_{n}(r)$). }
\label{singul02}
\end{figure}

\begin{figure}
\centering
\includegraphics[angle=0, scale=.8]{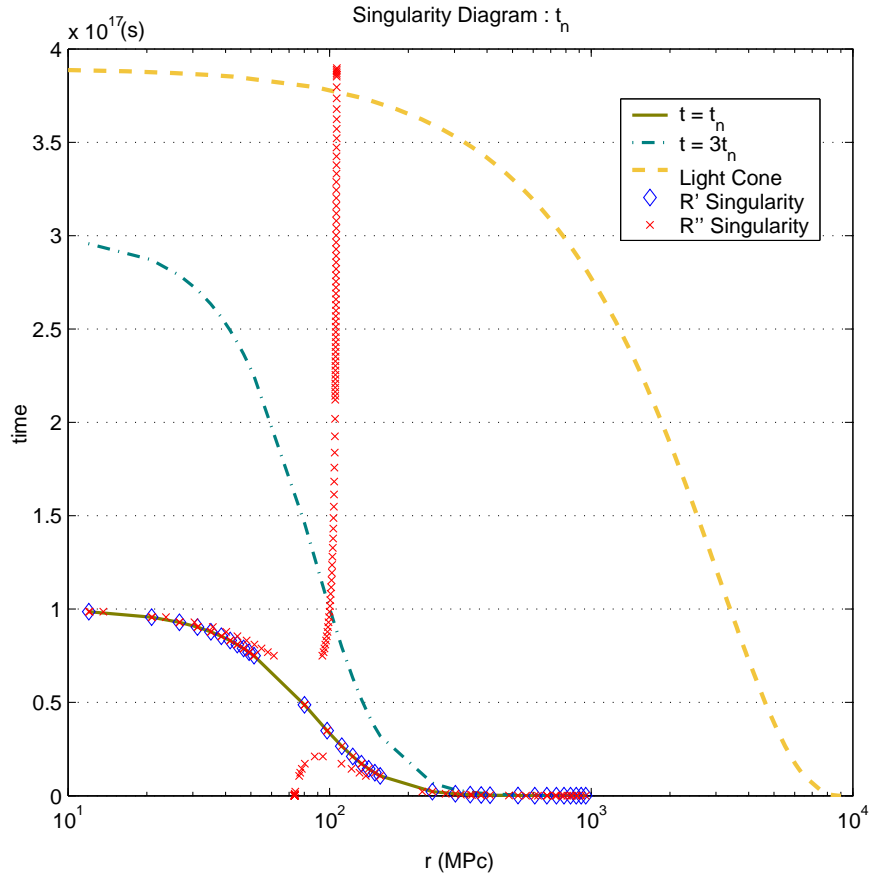}
\caption{Similar to Fig.3, for $t_{n}=\frac{\alpha}{r^4+r^2+1}\,\, ,
(\alpha=10^{17}~ s, n=1)$. All singularities are well inside the
curve $t=3t_{n}(r)$. The $\dot{R^{\prime}}$ singularity points
(corresponding to $t=t_{n}-\frac{1}{3}rt_{n}^{\prime}$) are not
shown because they almost lie completely on the curve for
$t=t_{n}(r)$.} \label{singul04}
\end{figure}

\begin{figure}
\centering
\includegraphics[angle=0, scale=.6]{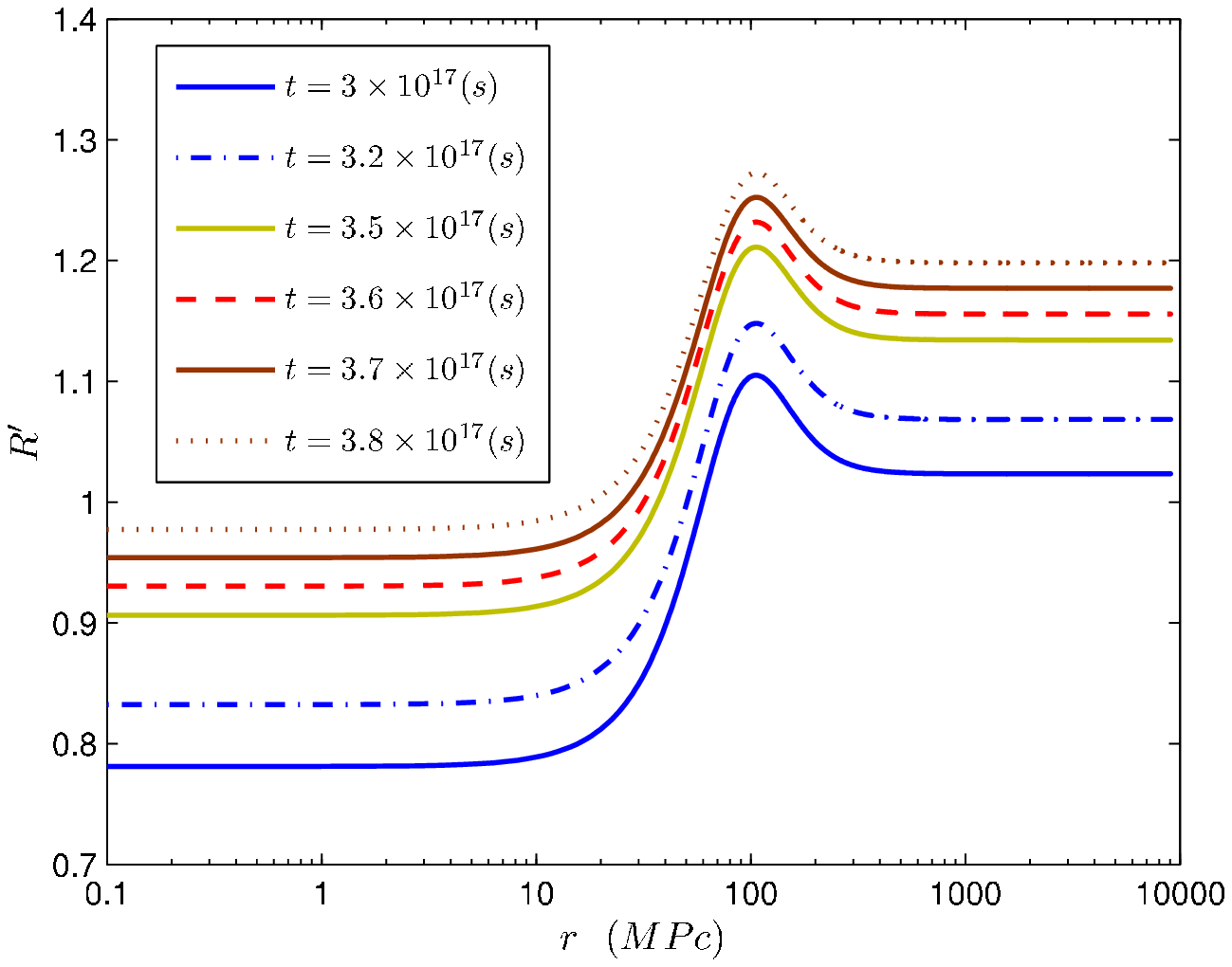}
\caption{ $R'$ is plotted versus $r$ for $t_{n}$ and different
constant times. Looking at $R'$ as an effective scale factor, it
shows that the scale of the universe increases with time, although
the rate of cosmic expansion is different at different places. }
\label{rp1}
\end{figure}

\begin{figure}
\centering
\includegraphics[angle=0, scale=.6]{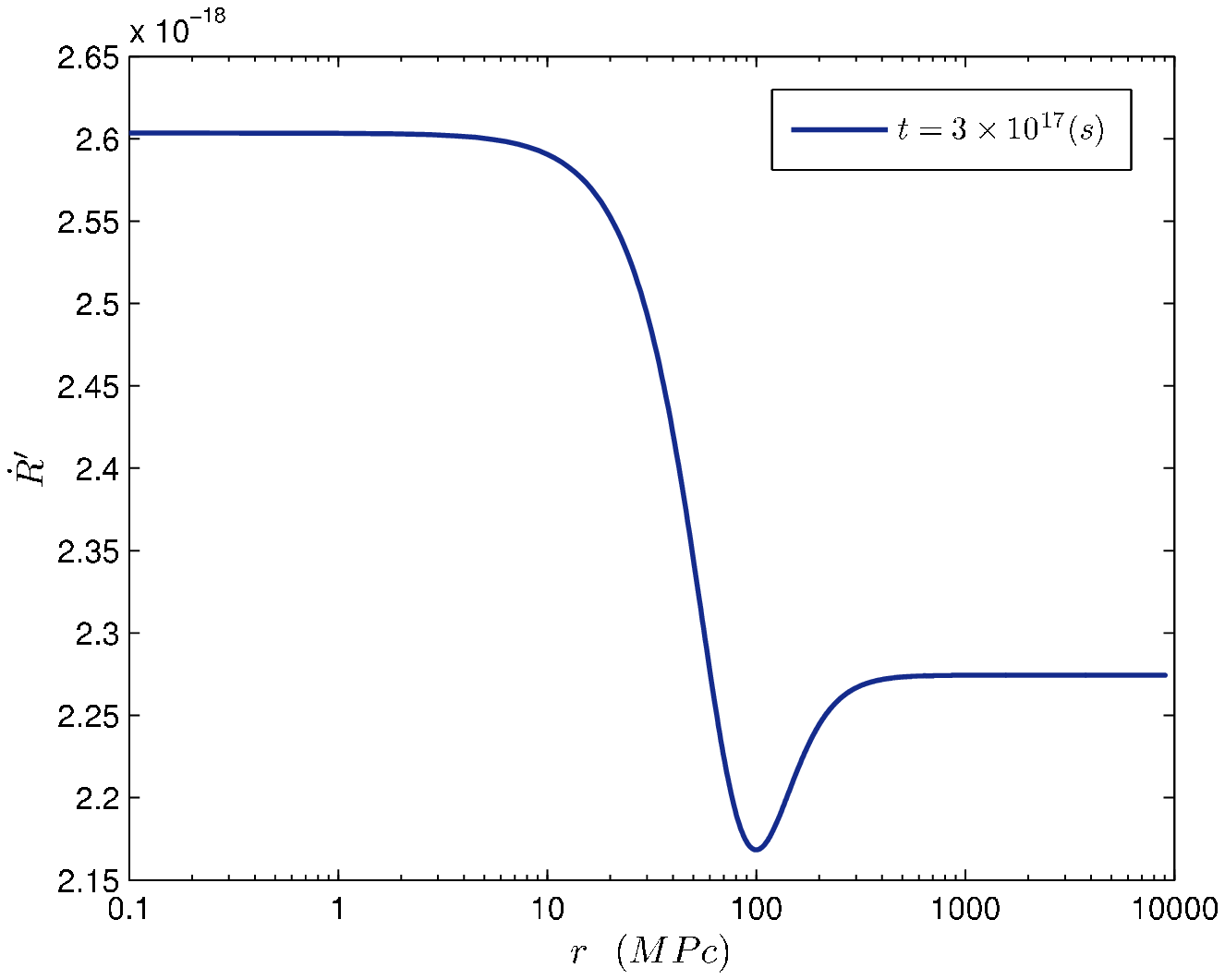}
\caption{ $\dot{R}'$ vs $r$ for $t_{n}$ and $t=3 \times 10^{17} ~s$.
No singularity is seen here.} \label{rpd1}
\end{figure}

\begin{figure}
\centering
\includegraphics[angle=0, scale=.6]{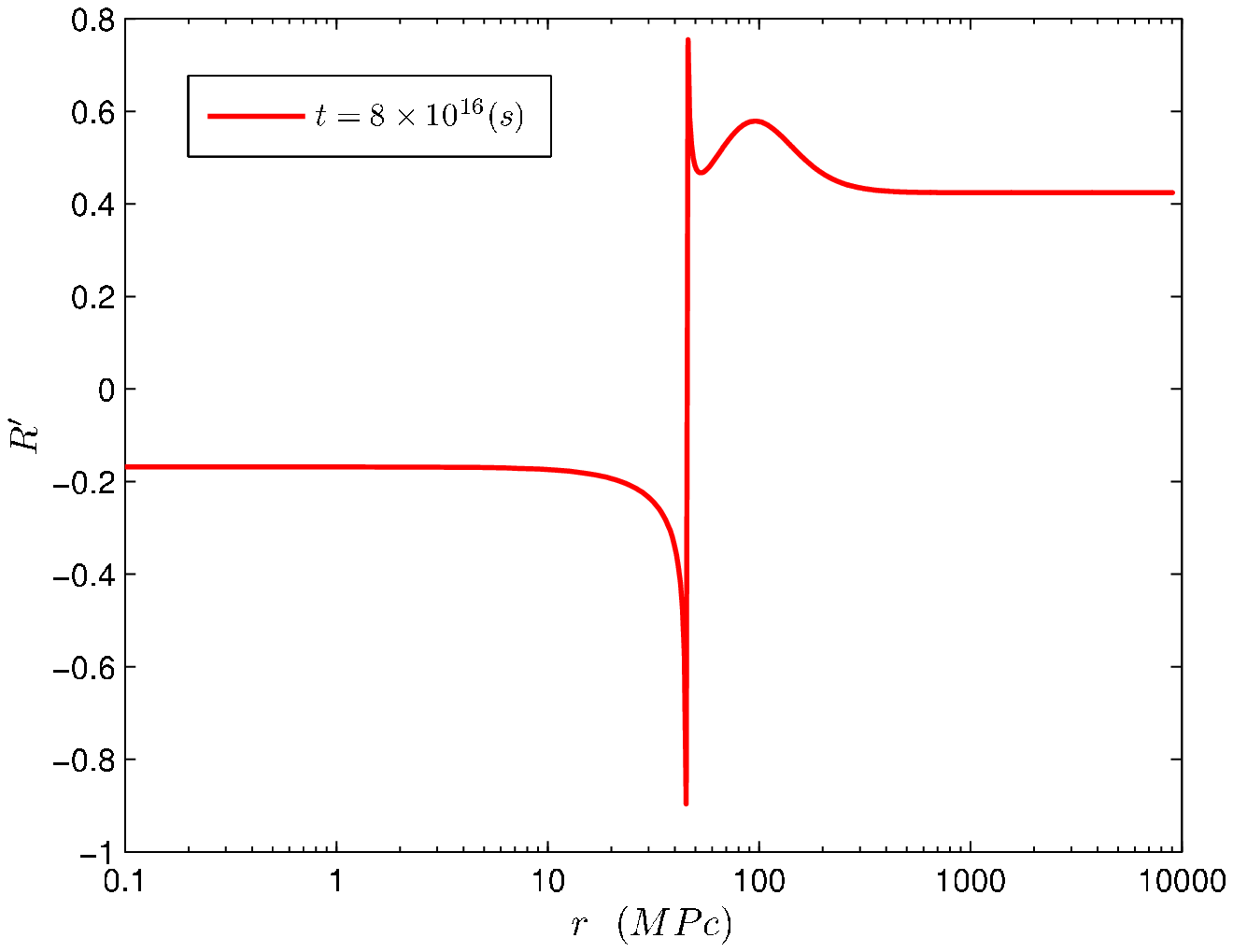}
\caption{ A typical singular behavior of $R'$ inside the light cone
for $t_{n}$ and $t=8 \times 10^{16}~s$. The bump after the singular
point corresponds to the vanishing of $R''$.} \label{rp2}
\end{figure}

\begin{figure}
\centering
\includegraphics[angle=0, scale=.6]{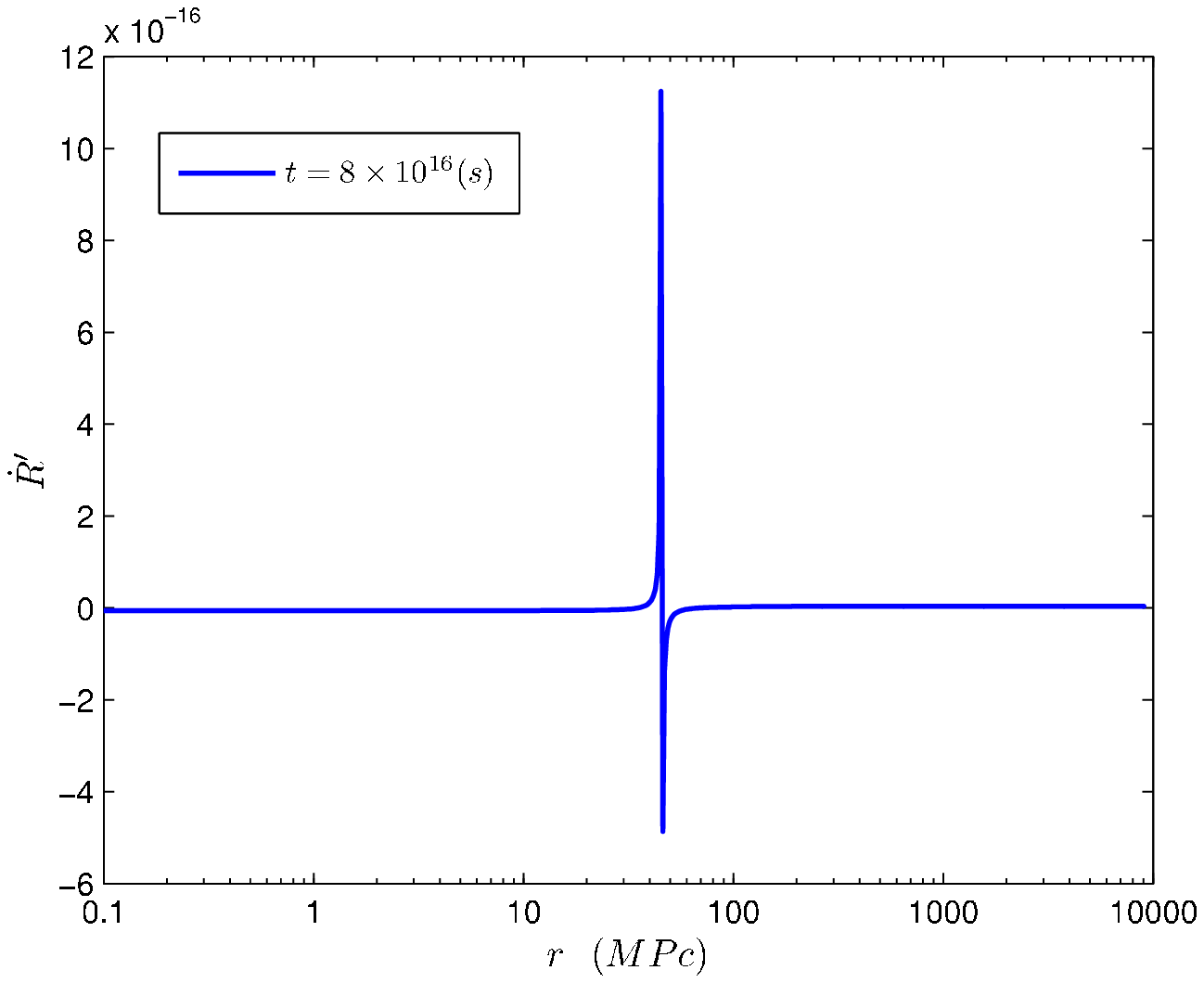}
\caption{A typical singular behavior of $\dot{R}'$ vs $r$ for
$t_{n}$ and constant $t=8 \times 10^{16} ~s$. } \label{rp3}
\end{figure}

\begin{figure}
\centering
\includegraphics[angle=0, scale=.5]{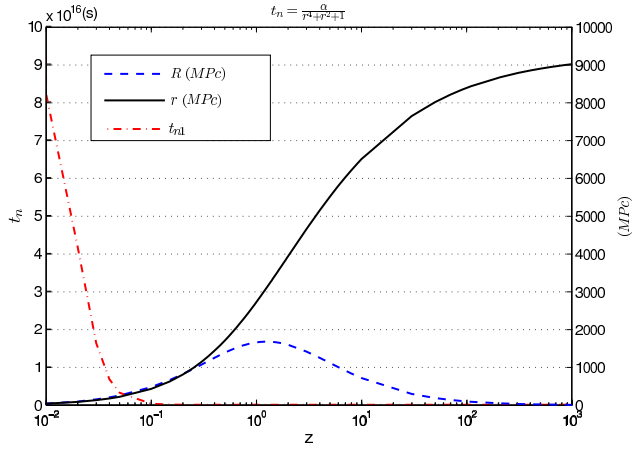}
\caption{ $t_{n}$, $R$, $r$ is drawn versus $z$. The bang time
($t_{n}$) is almost zero for $z>0.1$ . When we go back in time along
the light cone, the scale of the universe (blue curve) goes to zero
(i.e. $R \to 0$).} \label{tnRr}
\end{figure}

\begin{figure}
\centering
\includegraphics[angle=0, scale=.6]{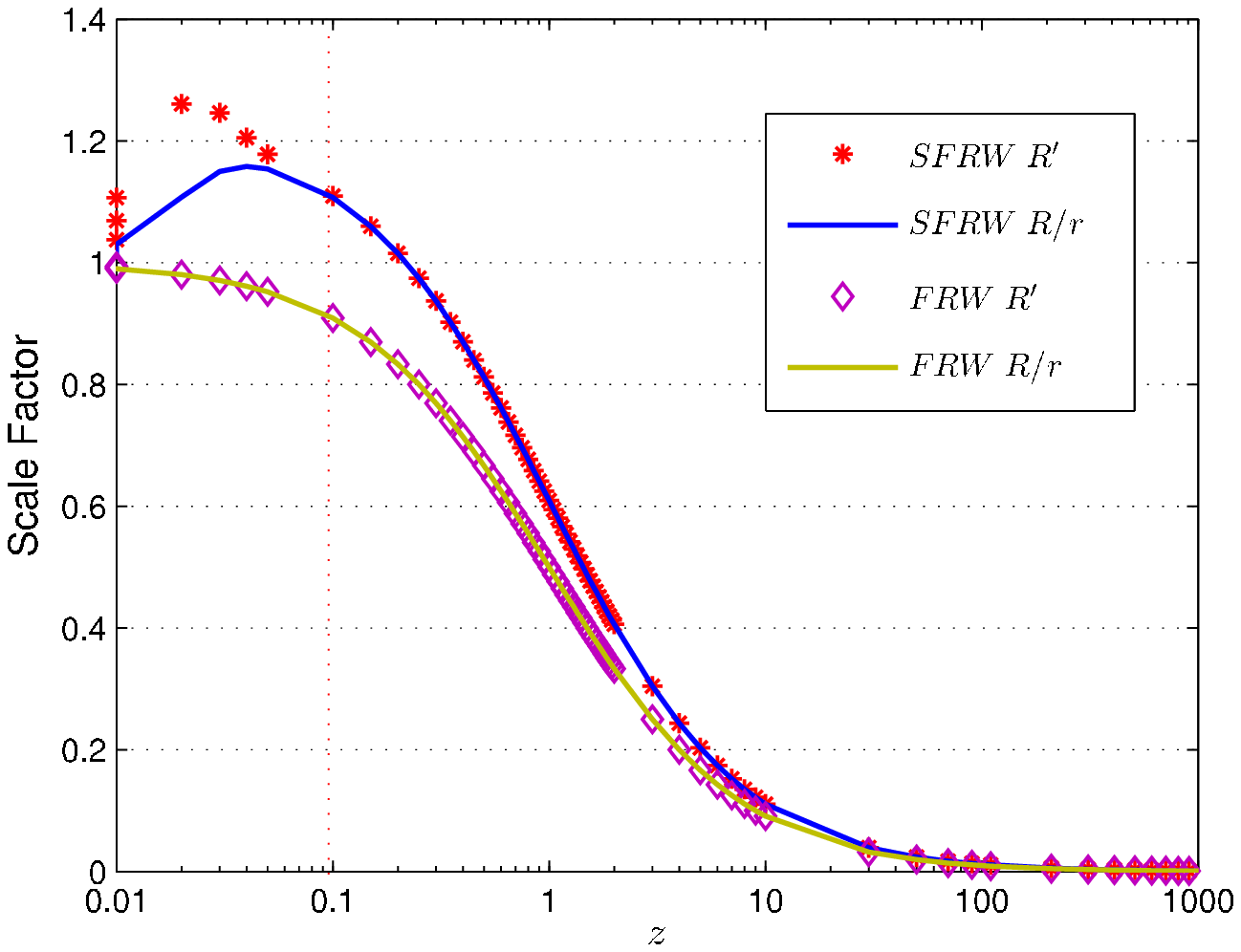}
\caption{ The SFRW effective scale factor $R/r$ as a function of $z$
for $t_{n}$ (blue curve). The red points show $R'$ as a function of
$z$. Both are almost the same for $z>0.1$. The effect of
inhomogeneities is more clear for $z<0.1$ where $R' \ne R/r $. The
green curve shows the scale factor for a CDM/FRW model ($\alpha=0$).
} \label{scalefactor}
\end{figure}

\begin{figure}
\centering
\includegraphics[angle=0, scale=.6]{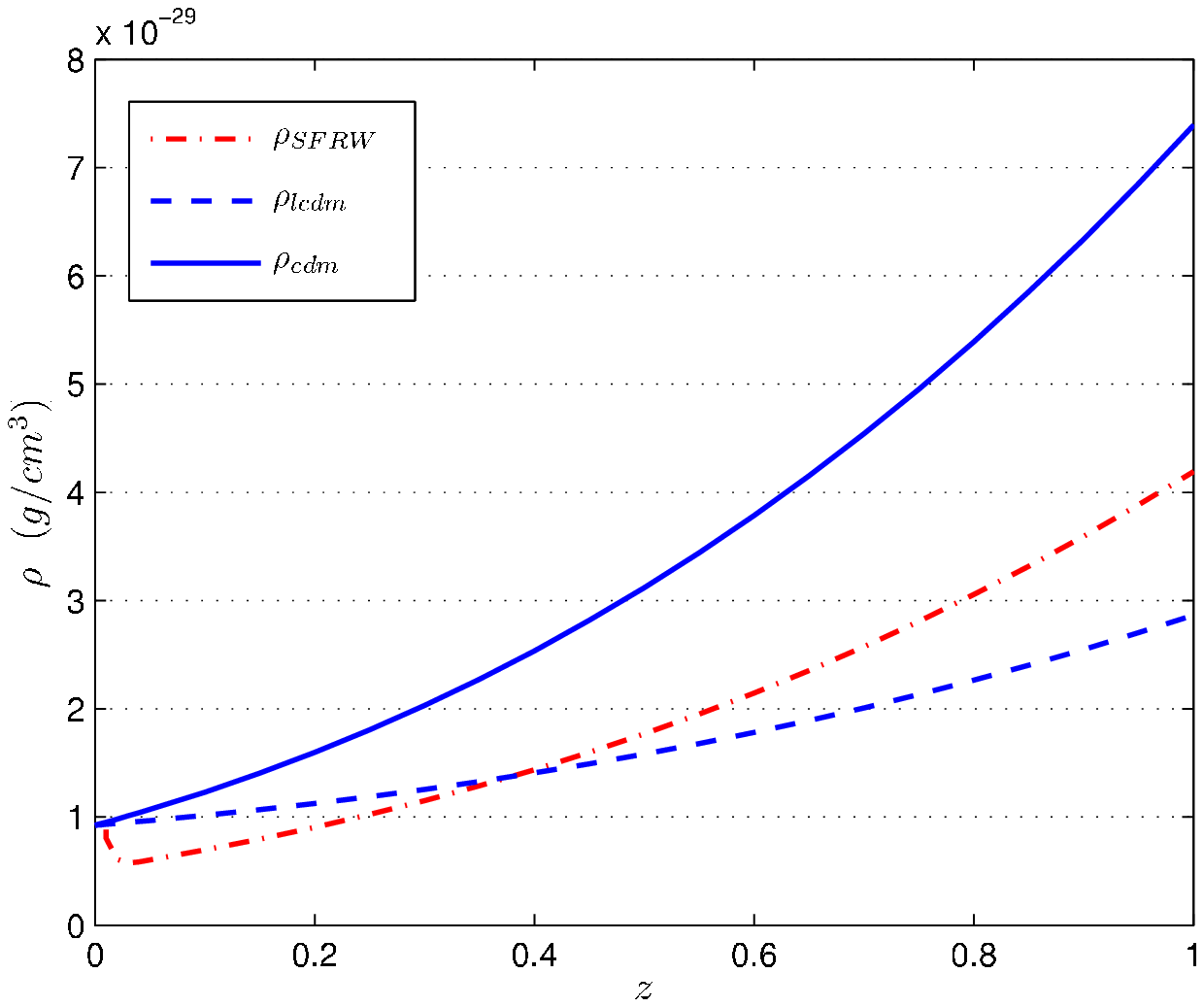}
\caption{Density $\rho$ as a function of $z$ for $t_{n}$ in low
redshift region} \label{rho1}
\end{figure}

\begin{figure}
\centering
\includegraphics[angle=0, scale=.6]{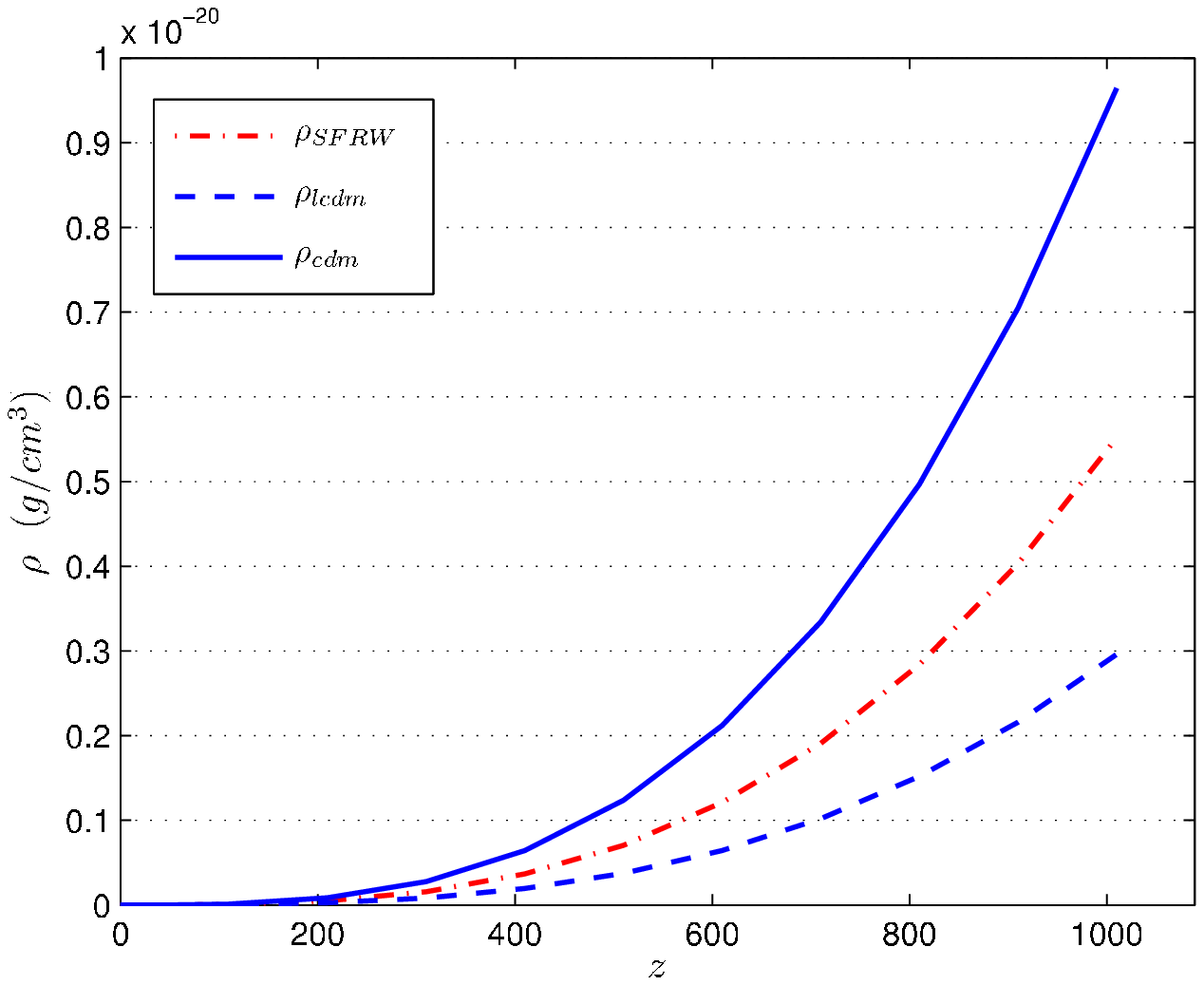}
\caption{Density $\rho$ as a function of $z$ for $t_{n}$ in high
redshift region} \label{rho2}
\end{figure}

\begin{figure}
\centering
\includegraphics[angle=0, scale=.8]{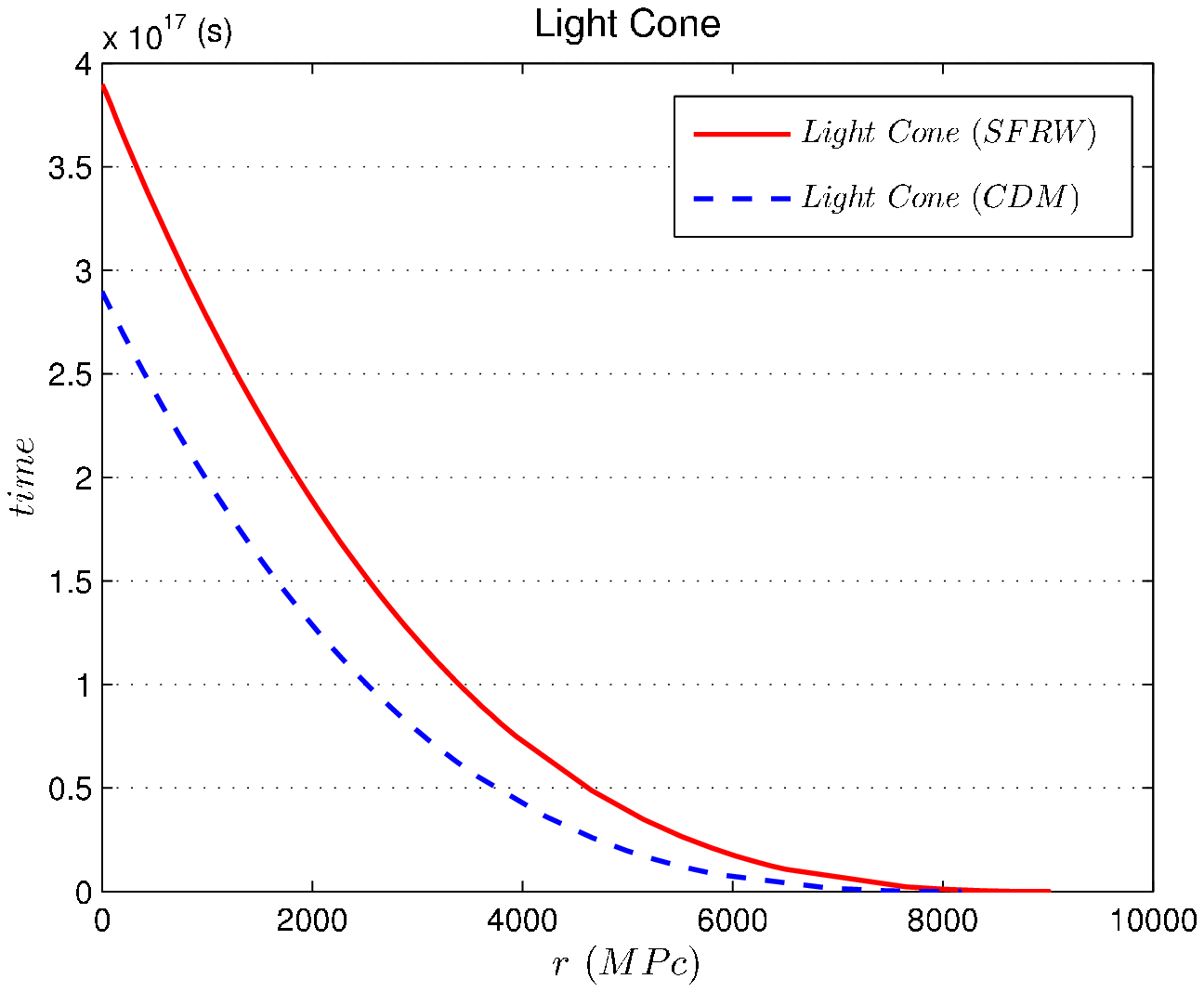}
\caption{ Comparison of light cones for both SFRW (LTB) and FRW
(CDM) models. The age of the Universe is greater in SFRW model by
the factor $4/3$. We got these results for $z<1100$ (after last
scattering time).} \label{lightcone}
\end{figure}

\begin{figure}
\centering
\includegraphics[angle=0, scale=.8]{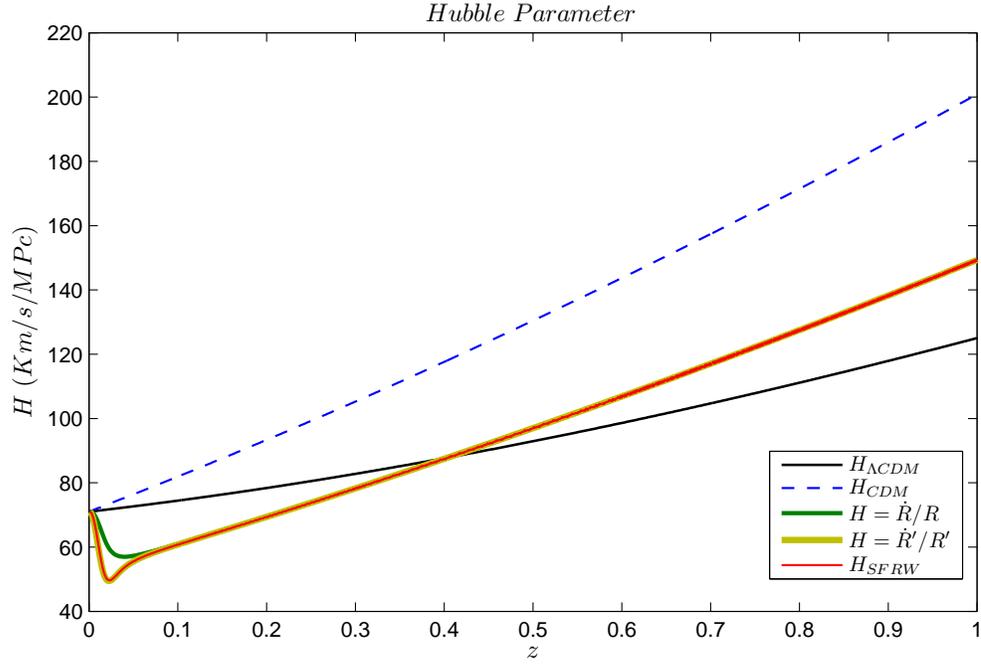}
\caption{ Using the luminosity distance and eq.\ref{h}, the
effective Hubble parameter is plotted. It shows also the Hubble
parameter for CDM and $\Lambda$CDM models. Comparison of the
effective Hubble parameter based on the luminosity distance
($H_{SFRW}$) and $ \dot{R}'/R'$ shows, for all values of redshift
parameter, these two definitions are the same.
($t_{n}=\frac{\alpha}{r^4+r^2+1},~n=1,~\alpha=10^{17}~s$)}\label{hubble}
\end{figure}

\begin{figure}
\centering
\includegraphics[angle=0, scale=.9]{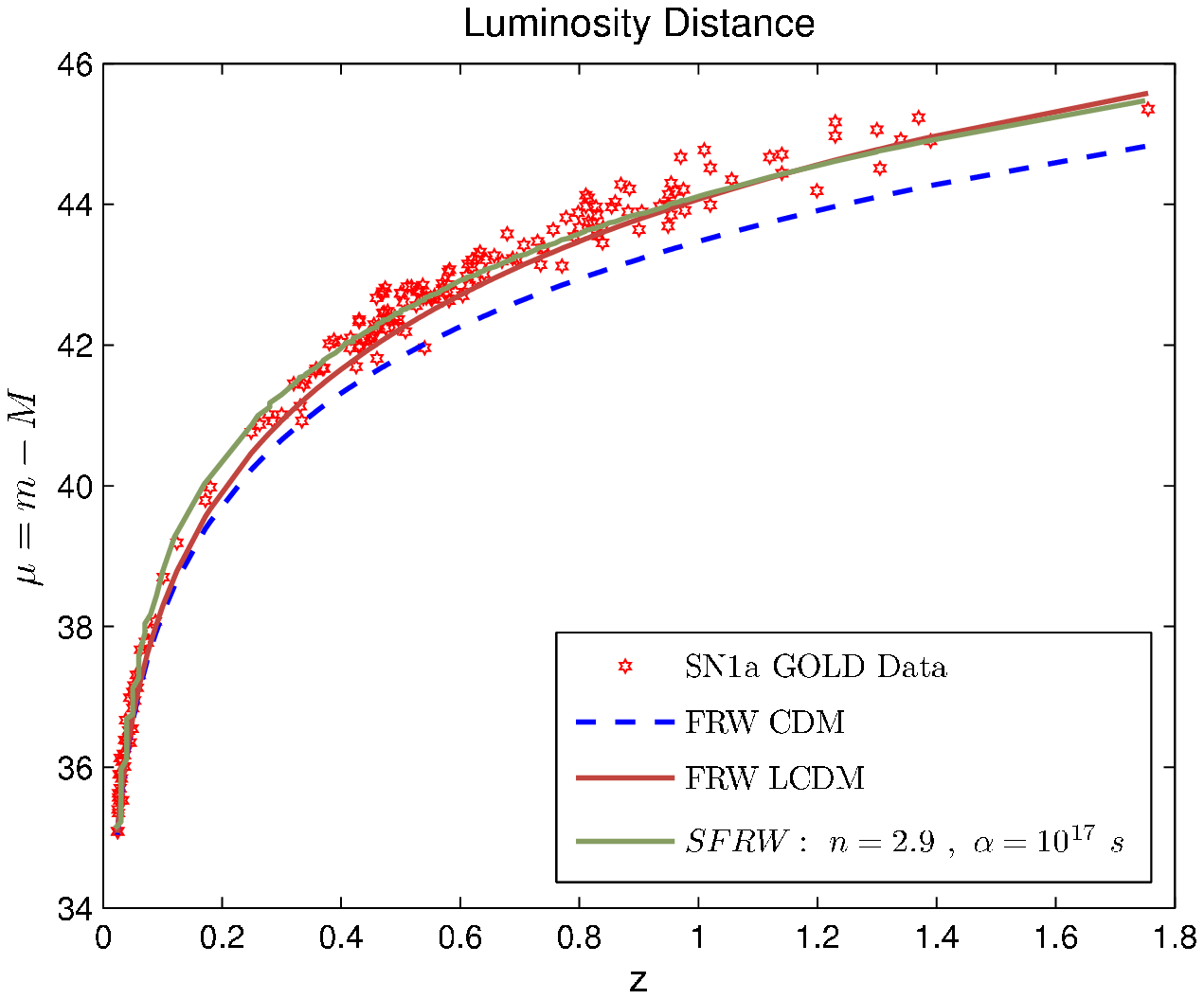}
\caption{ $\mu=m-M$ vs $z$ (redshift) for different models (LCDM,
CDM, $t_{n}$ SFRW).} \label{lumino}
\end{figure}

\begin{figure}
\centering
\includegraphics[angle=0, scale=.6]{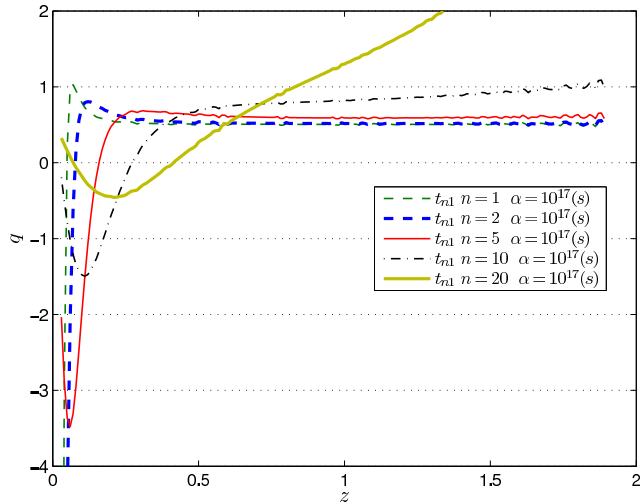}
\caption{The effective deceleration parameter for SFRW model
($t_{n1}=\frac{\alpha}{r^2+1}$) using the effective $H(z)$ and
eq.\ref{q}. This diagram clearly shows that, for small values of
$z$, $q$ is negative and therefore, the universe is effectively
accelerating at the present time. Increasing the parameter $n$, the
transition of acceleration to deceleration occurs in higher
redshifts. }\label{deceler}
\end{figure}

\begin{figure}
\centering
\includegraphics[angle=0, scale=.6]{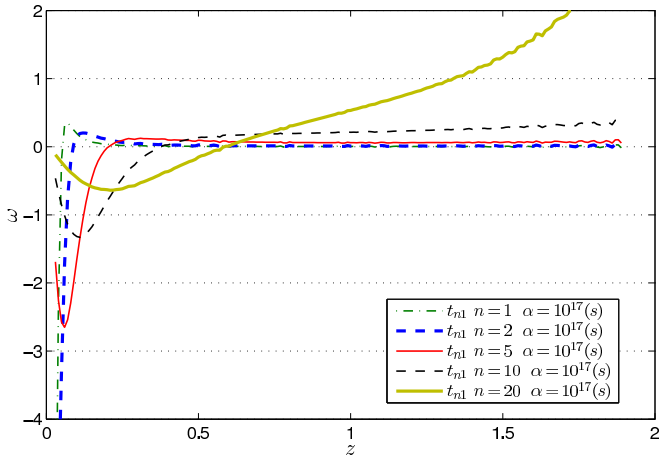}
\caption{The effective state parameter for SFRW model
($t_{n1}=\frac{\alpha}{r^2+1}$) using the effective $H(z)$ and
eq.\ref{w}. }\label{deceler01}
\end{figure}

\begin{figure}
\centering
\includegraphics[angle=0, scale=.6]{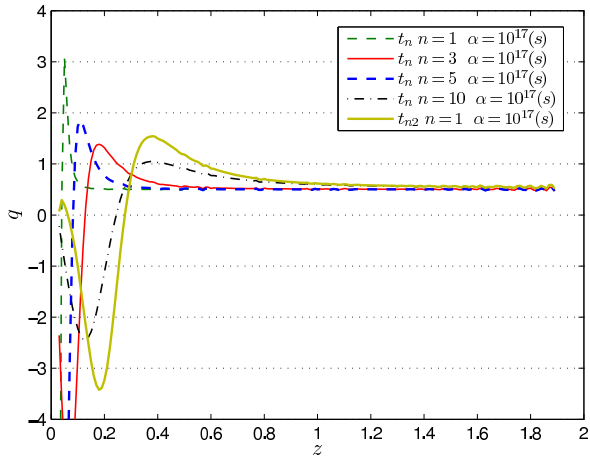}
\caption{ Similar to Fig.\ref{deceler}, using $t_{n} =
\frac{\alpha}{r^4+r^2+1}$ and $t_{n2} = \frac{\alpha}{Ar^4+Br^2+1}$
for $A=0.0001,~B=0.001$.}\label{deceler2}
\end{figure}

\begin{figure}
\centering
\includegraphics[angle=0, scale=.6]{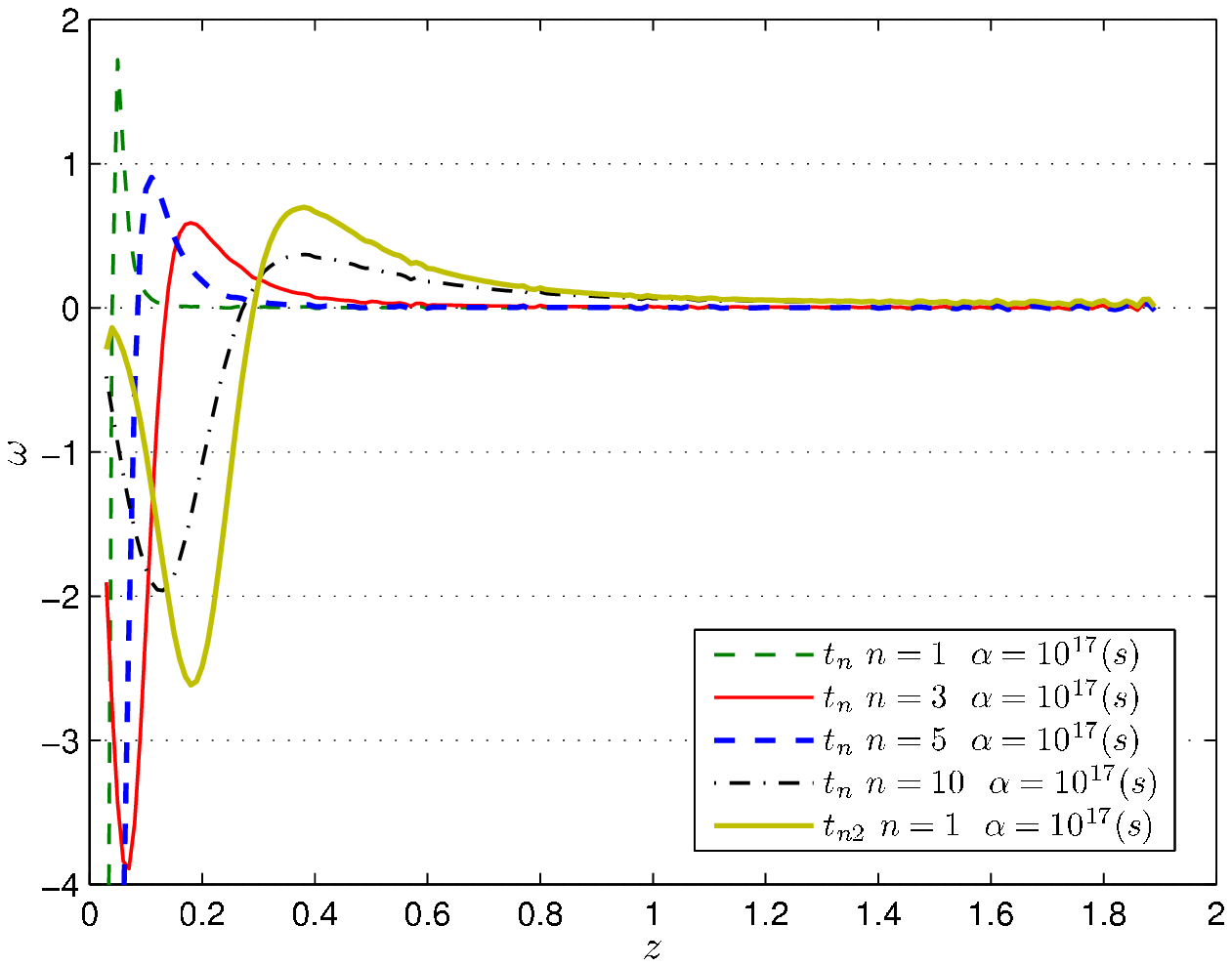}
\caption{ Similar to Fig.\ref{deceler01}, using $t_{n} =
\frac{\alpha}{r^4+r^2+1}$ and $t_{n2} =
\frac{\alpha}{Ar^4+Br^2+1}$.} \label{deceler02}
\end{figure}

\end{document}